\documentclass[preprint,showpacs,preprintnumbers,amsmath,amssymb]{revtex4}

\usepackage{color,vmargin,epsfig,rotating,graphicx,subfigure}
\usepackage{amsmath,amssymb,amscd}
\usepackage[latin1]{inputenc}
\usepackage{dsfont}

\setpapersize{A4}
\renewcommand{\Im}{{\rm Im}}

\newcommand{\ri}{{\rm i}}
\newcommand{\re}{{\rm e}}
\newcommand{\rd}{{\rm d}}
\newcommand{\kb}{k_{\rm B}}

\newcommand{\rP}{{\rm P}}
\newcommand{\rS}{{\rm S}}
\newcommand{\rR}{{\rm R}}
\newcommand{\rE}{{\rm E}}
\newcommand{\rH}{{\rm H}}
\newcommand{\Tr}{{\rm Tr}}

\newcommand{\GE}[2]{\mathds{G}^{\rm E}(\mathbf{#1},\mathbf{#2})}

\newcommand{\GER}[2]{\mathds{G}^{\rm E}_{\rm R}(\mathbf{#1},\mathbf{#2})}
\newcommand{\GHR}[2]{\mathds{G}^{\rm H}_{\rm R}(\mathbf{#1},\mathbf{#2})}

\begin{document}

\title{Near-field radiative heat transfer for structured surfaces}

\author{Svend-Age Biehs, Oliver Huth, Felix R\"uting}

\affiliation{Institut f\"ur Physik, Carl von Ossietzky Universit\"at,
        D-26111 Oldenburg, Germany}

\date{July 04, 2008}

\preprint{Physical Review B, in press}


\begin{abstract}
We apply an analytical approach for determining the near-field
radiative heat transfer between a metallic nanosphere and a
planar semi-infinite medium with some given surface structure.
This approach is based on a perturbative expansion, and evaluated
to first order in the surface profile. With the help of numerical
results obtained for some simple model geometries we discuss typical
signatures that should be obtainable with a near-field scanning
thermal microscope operated in either constant-height or
constant-distance mode.
\end{abstract}

\maketitle

\section{Introduction}
\label{Sec1}

Recent progress in the fabrication of a near-field scanning thermal microscope (NSThM) enables one to
measure the radiative heat transfer between a cooled sample and a hot probe directly in the near-field regime, i.e.\
for distances in the nanometer range~\cite{KittelEtAl05,WishnathEtAl08}. 
Due to thermally excited evanescent waves, in this regime one expects an energy transfer several orders of magnitude greater 
than the black body value~\cite{PolderVanHove71}. For estimating the heat current in such a device several theoretical models 
are available, which describe the probe as a dielectric sphere, and the sample as a semi-infinite dielectric body with a flat 
surface~\cite{Dorofeyev98,JPendry99,MuletEtAl01,A.I.VolokitinB.N.J.Persson01,ChapuisEtAl08,Dorofeyev08,DedkovKyasov07}. These
models can now be tested against the data provided by the NSThM.   
       
\begin{figure}[Hhbt]
  \centering
  \begin{minipage}[t]{0.9\textwidth}
    \epsfig{file=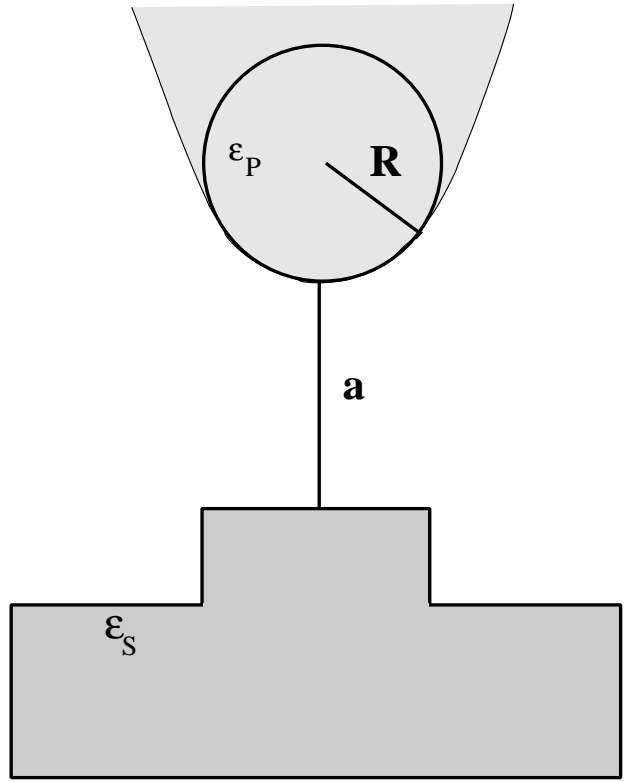, width = 0.6\textwidth}
    \caption{Sketch of a probe-sample configuration with a square pad on a flat surface.}
    \label{Fig:Configuration}
  \end{minipage}
\end{figure}

In the literature, the near-field radiative heat transfer between a sphere and a 
structured surface has not been studied so far, although such a geometry is of considerable practical relevance and theoretical 
interest. Therefore, in the present paper we analyze the near-field radiative heat transfer 
between a spherical probe and planar samples with surface structures such as depicted in 
Fig.~\ref{Fig:Configuration}. In particular, we discuss numerical results obtained
for a planar surface structured by an infinite bar and a square pad, respectively.    

In this work, we use the general formulation of the near-field radiative heat transfer between a probe
described as a spherical metallic nanoparticle within the dipole approximation and a second material 
as developed in Refs.~\cite{DedkovKyasov07,Dorofeyev08,ChapuisEtAl08}, which is based on 
Rytov's fluctuational electrodynamics~\cite{RytovEtAl89}. This formulation allows one to take the material's properties of the 
probe into account in terms of its electric and magnetic dipole moments, and the properties of the sample material, which is 
assumed to be a semi-infinite body with a given surface structure, through the local density of states~\cite{JoulainEtAl03} above 
that medium. Since the dipole moments of the probe or nanoparticle are known it remains to calculate the local
density of states (LDOS) above the sample material. This is done within a perturbative 
approximation employing the Ewald-Oseen extinction theorem, as described in detail in Ref.~\cite{G.S.Agarwal1977}. 
It should be mentioned that our work is closely related to Ref.~\cite{C.HenkelV.Sandoghdar98}, where the
changes of linewidth and the lineshift for a molecule near a structured surface were calculated, since the electric
LDOS can in principle be read off from the Green's function calculated therein. However, in our work not only
the electric but also the magnetic fields are determined, since it is known~\cite{YuMartynenko05,ChapuisEtAl08} that for 
metallic nanoparticles these magnetic fields can cause a heat transfer much greater than that due to the electric fields,
as a result of the induction of Foucault's currents.   

This paper is organized as follows: In Sec.~II we introduce the dipole
model for the radiative heat transfer between a metallic nanoparticle
and an arbitrary second material, and outline the strategy of the
following calculations. In order to determine the electromagnetic LDOS,
we deduce the appropriate dyadic Green's functions in Sec.~III by
means of a perturbative expansion due to Agarwal~\cite{G.S.Agarwal1977}, which we
terminate after the first order. Within this approach, the LDOS itself
is calculated and numerically evaluated for different surface profiles
in Sec.~IV, where we also deduce tentative criteria justifying the
restriction to the lowest-order contributions. In Sec.~V we then
present some numerical results for the near-field heat transfer
between a metallic nanosphere and a structured surface, and predict
signatures that should be observable with a NSThM operated in either
constant-height or constant-distance mode.

\section{Heat transfer between a Nanoparticle and a second medium}
\label{Sec2}

It has recently been shown~\cite{DedkovKyasov07,ChapuisEtAl08,Dorofeyev08} that the near-field 
radiative heat transfer between a metallic nanoparticle
with radius $R$ and temperature $T_\rP$ and a dielectric material with 
temperature $T_\rS$ can be described within the dipole model for particle-sample distances $a \gg R$. In this model
the electric polarisability $\alpha_\rP$ and the magnetic polarisability $\mu_\rP$ of the particle are given by~\cite{Landau60}
\begin{equation}
  \alpha_{\rm P}(\omega,R) = 4 \pi R^3 \frac{\epsilon_{\rm P}(\omega)-1}{\epsilon_{\rm P}(\omega)+2}
  \label{Eq:alpha_E}
\end{equation}
and
\begin{equation}
    \mu_{\rm P}(\omega,R) = -\frac{R^3}{2}\biggl[1 - 3\frac{d_{\rm s}'^2(\omega)}{R^2} +
                                     3 \frac{d_{\rm s}'(\omega)}{R}
                                     \cot\biggl( \frac{R}{d_{\rm s}'(\omega)} \biggr)  \biggr]
  \label{Eq:alpha_H}
\end{equation}
with the permittivity of the probe $\epsilon_{\rm P}(\omega)$ and $d_{\rm s}'(\omega) := c (\omega \sqrt{\epsilon_{\rm P}(\omega) - 1})^{-1}$; as usual, $c$ denotes the velocity of light in vacuum. Assuming that the two bodies are in local thermal 
equilibrium, the mean energy rate $P$ flowing from the hot to the cold body is given by the relation
\begin{equation}
  P = \int_0^\infty\!\!\rd \omega\, \bigl[ \Theta(\omega,T_\rP) - \Theta(\omega,T_\rS) \bigr] 2 \omega \bigl( 
      \alpha_\rP''(\omega,R) D^\rE(\omega,\mathbf{r}) + \mu_\rP''(\omega,R) D^\rH(\omega,\mathbf{r})\bigr).
\label{Eq:DipolModel}
\end{equation}
Here, the sign of the mean energy rate $P$ is determined by the difference of the Bose-Einstein functions
$\Theta(\omega,T):= \hbar \omega (\exp(\hbar \omega \beta) - 1)^{-1}$ with the  
temperature $\beta^{-1} := \kb T$, so that the sign is positive for $T_\rP > T_\rS$ and negative otherwise. 
The material's properties of the nanoparticle are taken into account in Eq.~(\ref{Eq:DipolModel}) 
by means of the imaginary part (as indicated by the double prime) of the electric and magnetic 
polarisabilities, $\alpha_\rP''(\omega,R)$ and $\mu_\rP''(\omega,R)$, whereas the material's and geometrical  
properties of the second medium enter into this expression via the electric and magnetic local density of states 
$D^\rE(\omega,\mathbf{r})$ and $D^\rH(\omega,\mathbf{r})$ at the point $\mathbf{r}$ above this medium, where the probe 
is located. Therefore, the thermal radiative heat transfer between a nanoparticle and an arbitrary second medium can be calculated if the
electromagnetic local density of states above that medium is known. We point out that in the opposite limit 
$a \ll R$ the dipole model discussed here is not valid, whereas the so called ``proximity approximation'' should 
prove to be useful then~\cite{VolokitinPersson2007}. 

Now the local density of states can be calculated with the help of the relations~\cite{JoulainEtAl05}
\begin{equation}
  D^\rE(\omega,\mathbf{r}) = \frac{\omega}{\pi c^2} \epsilon_{\rm S}'' (\omega) \int\!\! \rd^3 r'\, \Tr\biggl[ \GE{r}{r'} \GE{r}{r'}^\dagger \biggr]
  \label{Eq:non_equi_electric_LDOS}
\end{equation}
and
\begin{equation}
  D^\rH(\omega,\mathbf{r}) = \frac{\omega}{\pi c^2} \frac{\epsilon_{\rm S}'' (\omega)}{\omega^2 \mu_0^2} \int\!\! \rd^3 r'\, \Tr\biggl[ \nabla\times\GE{r}{r'}(\nabla\times\GE{r}{r'})^\dagger \biggr],
  \label{Eq:non_equi_magnetic_LDOS}
\end{equation} 
in the case of local equilibrium inside a heat-radiating local medium surrounded by vacuum. Here
$\mu_0$ is the magnetic permeability of the vacuum, and $\epsilon_{\rm S}(\omega)$ is the relative permittivity of the 
material considered. These relations can in principle be evaluated by determining the electric dyadic Green's function
$\GE{r}{r'}$ with the source points $\mathbf{r}'$ inside the medium and the observation
point $\mathbf{r}$ outside the medium, implementing the tensor product with its hermitian conjugate and integrating over
the volume of the medium. Since, here only a local equilibrium inside the medium is assumed, one can use these relations
to investigate for example the heat transfer between bodies kept at different temperatures. 

Considering a medium surrounded by vacuum, the non-equilibrium expressions in Eqs.~(\ref{Eq:non_equi_electric_LDOS}) and
(\ref{Eq:non_equi_magnetic_LDOS}) can be decomposed into an evanescent and a propagating part by transforming the 
volume integral into a surface integral~\cite{VolokitinPersson2007}. 
This evanescent part of the local density of states coincides with the equilibrium expression, since the 
evanescent modes are bound to the surface of the medium and are therefore not 
relevant for preserving a global equilibrium situation. Thus, the evanescent part of the local density of states 
above a material surrounded by vacuum can also be calculated by means of the equilibrium 
expressions~\cite{JoulainEtAl05,VolokitinPersson2007}
\begin{equation}
  D^\rE(\omega,\mathbf{r}) = \frac{\omega}{\pi c^2} \Im \,\Tr\, \GER{r}{r}
  \label{Eq:equi_electric_LDOS}
\end{equation}
and
\begin{equation}
  D^\rH(\omega,\mathbf{r}) = \frac{\omega}{\pi c^2} \Im \, \Tr\, \GHR{r}{r},
  \label{Eq:equi_magnetic_LDOS}
\end{equation}
which state that the electric and magnetic local density of states at the point $\mathbf{r}$ is given by 
the imaginary part of the trace of the 
renormalised Green's functions $\GER{r}{r}$ and $\GHR{r}{r}$, where the renormalisation procedure is defined as~\cite{Landau92}
\begin{equation}
  \mathds{G}^{\rE/\rH}_{\rm R} (\mathbf{r,r}) = \lim_{\mathbf{r} \rightarrow \mathbf{r}'} \biggl[\mathds{G}^{\rE/\rH} (\mathbf{r,r'}) - \mathds{G}^{\rE/\rH}_0 (\mathbf{r,r'})  \biggr] 
\label{Eq:renormalisation}
\end{equation}
with $\mathds{G}^{\rE/\rH}_0 (\mathbf{r,r'})$ denoting the Green's function of the vacuum. Since the Green's function with 
the observation and source point located above the medium consists
of an incident and a reflected part, with the incident part coinciding with $\mathds{G}^{\rE/\rH}_0 (\mathbf{r,r'})$, the renormalised  
dyadic Green's function coincides with the reflected Green's function, so that the index ``R'' can be understood as both ``renormalised'' and ``reflected''.

Obviously, the relations in Eqs.\ (\ref{Eq:equi_electric_LDOS}) and (\ref{Eq:equi_magnetic_LDOS}) are much easier to evaluate than the 
expressions in Eqs.\ (\ref{Eq:non_equi_electric_LDOS}) and (\ref{Eq:non_equi_magnetic_LDOS}), 
giving reliable results for such distances above the material, at which the evanescent modes dominate the local density of states.
Since we are especially interested in the evanescent regime, we focus on the equilibrium
relations in order to determine the radiative heat transfer between a nanoparticle and a structured surface, keeping in mind that
the results hold in the evanescent regime only. Thus, it is necessary
to calculate the reflected electric and magnetic Green's function with observation points $\mathbf{r}$ and source points 
$\mathbf{r}'$ located outside the material of interest. For an
electric source current $\mathbf{j}_{e0}$ and a magnetic source current $\mathbf{j}_{m0}$ located at 
the point $\mathbf{r}'$, the reflected fields and reflected dyadic Green's functions are connected by the relations
\begin{equation}
  \mathbf{E}_\rR(\omega,\mathbf{r}) = \ri \omega \mu_0 \GER{r}{r'} \mathbf{j}_{e0}
\label{Eq:E_field_greens_function}
\end{equation}
and
\begin{equation}
  \mathbf{H}_\rR(\omega,\mathbf{r}) = \ri \omega \mu_0 \GHR{r}{r'} \mathbf{j}_{m0}.
\label{Eq:H_field_greens_function}
\end{equation}
Furthermore, in vacuum the electric and magnetic dyadic Green's functions are related by~\cite{Felsen94}   
\begin{equation}
  \GHR{r}{r'} = - \frac{1}{k_0^2} \nabla\times\GER{r}{r'}\times\nabla'
\label{Eq:Green_electric_magnetic}
\end{equation}
with $k_0 := \omega/c$. 

In the following, we will therefore perturbatively evaluate the reflected electric 
field generated by an electric current $\mathbf{j}_{e0}$ located at the source point $\mathbf{r}'$
above a semi-infinite medium with a structured surface,
determine the electric dyadic Green's function by means of Eq.\ (\ref{Eq:E_field_greens_function}), and from this result  
deduce the magnetic dyadic Green's function from Eq.\ (\ref{Eq:Green_electric_magnetic}). Finally, we calculate
the local densities of states above the structured surface with Eqs.\ (\ref{Eq:equi_electric_LDOS}) and (\ref{Eq:equi_magnetic_LDOS}).
These densities, in their turn, then allow us to determine the near-field radiative heat transfer between 
a metallic nanoparticle and a semi-ininite medium with a structured surface.

\section{Green's function above a structured surface}
\label{Sec3}

\begin{figure}[Hhbt]
  \centering
  \begin{minipage}[t]{0.75\textwidth}
    \centering
    \epsfig{file=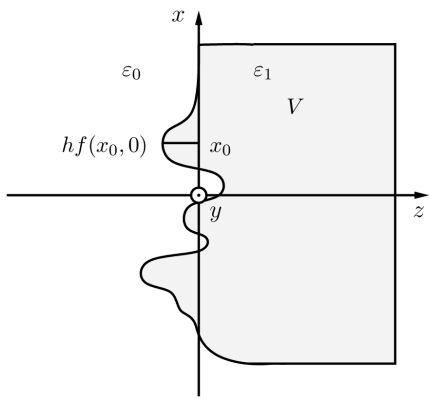, width = 0.6\textwidth}
    \caption{Sketch of an example structure, showing a particular value of the surface profile $h f(x,y)$
             at $x = x_0$ and $y = 0$.}
    \label{Fig:profil_geometry}
  \end{minipage}
\end{figure}

In order to calculate the electromagnetic fields above a structured surface, we assume that the 
surface profile is given by an expression $h f(x,y)$, where the dimensionless function $f(x,y)$
varies between zero and unity, and $h$ is the characteristic scale of the profile variation, as sketched 
in Fig.~\ref{Fig:profil_geometry}. Moreover, 
we assume that $h$ be small compared to all other relevant length scales of the problem, so that
we can apply a perturbation expansion. This approach has been worked out in great detail by 
Agarwal~\cite{G.S.Agarwal1977}, so that it suffices here to mention only those elements that are indispensable to follow
our line of reasoning.  

Within this approach the so-called Ewald-Oseen extinction theorem \cite{PattanayakWolf72,M.Vesperinas06} is employed, 
allowing us to restate the boundary conditions of the electromagnetic fields in the given geometry as integral equations.
For the case of a non-magnetic, isotropic, local, and linear material this theorem states that for 
observation points $\mathbf{r}$ outside that material, i.e. $\mathbf{r} \notin V$ 
(see Fig.\ \ref{Fig:profil_geometry}), one has  
\begin{equation}
  \begin{split}
    \mathbf{E}(\mathbf{r}) &= \mathbf{E}_{\rm I}(\mathbf{r}) + \mathbf{E}_{\rm R}(\mathbf{r})\\
                     &=\mathbf{E}_{\rm I} + \frac{1}{k_0^2} \boldsymbol{\nabla}\times\boldsymbol{\nabla}
		     \times \int_{\partial V} \rd S' \left[\mathbf{E}_{\rm T}(\mathbf{r}')
			\frac{\partial g(\mathbf{r}-\mathbf{r}')}{\partial \mathbf{n}'} -
			g(\mathbf{r}-\mathbf{r}') \frac{\partial \mathbf{E}_{\rm T}(\mathbf{r}')}
			 {\partial \mathbf{n}'}\right].        
  \end{split}\label{Eq:Ewald1}
\end{equation}
Here, the field $\mathbf{E}(\mathbf{r})$ outside the medium is simply the sum of the incident field
$\mathbf{E}_{\rm I}(\mathbf{r})$ and the reflected field $\mathbf{E}_{\rm R}(\mathbf{r})$. The latter
is described by the surface integral in Eq.~(\ref{Eq:Ewald1}), so that the reflected field
$\mathbf{E}_{\rm R}(\mathbf{r})$ can be calculated by means of the free Green's function
\begin{equation}
  g(\mathbf{r}-\mathbf{r}') = \frac{\re^{\ri k_0 |\mathbf{r}-\mathbf{r}'|}}{4 \pi |\mathbf{r}-\mathbf{r}'|}
\end{equation}
and the transmitted field $\mathbf{E}_{\rm T}(\mathbf{r})$. In addition, $\frac{\partial}{\partial \mathbf{n}'}$
symbolizes the normal derivative, taken in the direction of the unit normal of the surface,
\begin{equation}
  \mathbf{n}' = - \frac{\mathbf{e}_{\rm z} + h \boldsymbol{\nabla}'_\parallel f(x',y')}{\sqrt{1 + h^2 
                |\boldsymbol{\nabla}'_\parallel f(x',y')|^2}},
\end{equation}
with $\boldsymbol{\nabla}'_\parallel := (\partial_{x'},\partial_{y'},0)^t$ and $\mathbf{e}_z$ the unit vector in $z$-direction.

On the other hand, for observation points $\mathbf{r}$ within the medium, i.e.\ $\mathbf{r} \in V$, the Ewald-Oseen 
theorem gives 
\begin{equation}
    0 = \mathbf{E}_{\rm I} + \frac{1}{k_0^2} \boldsymbol{\nabla}\times\boldsymbol{\nabla}
		     \times \int_{\partial V} \rd S' \left[\mathbf{E}_{\rm T}(\mathbf{r}')
			\frac{\partial g(\mathbf{r}-\mathbf{r}')}{\partial \mathbf{n}'} -
			g(\mathbf{r}-\mathbf{r}') \frac{\partial \mathbf{E}_{\rm T}(\mathbf{r}')}
			 {\partial \mathbf{n}'}\right].        
  \label{Eq:Ewald2}
\end{equation}
In fact, the surface integrals in Eqs.~(\ref{Eq:Ewald1}) and (\ref{Eq:Ewald2}) have the same structure, but
one has to keep in mind that $\mathbf{r}\notin V$ in Eq.~(\ref{Eq:Ewald1}), whereas $\mathbf{r} \in V$
in Eq.~(\ref{Eq:Ewald2}), so that both integrals give different results. 
By means of Eq.~(\ref{Eq:Ewald2}) the transmitted field $\mathbf{E}_{\rm T}(\mathbf{r})$ can be computed if 
the incident field $\mathbf{E}_{\rm I}(\mathbf{r})$ is given, so that the
sought-after reflected field $\mathbf{E}_{\rm R}(\mathbf{r})$ can then be determined with the
surface integral in Eq.~(\ref{Eq:Ewald1}).

Now the transmitted and reflected fields in Eqs.~(\ref{Eq:Ewald1}) and (\ref{Eq:Ewald2}) 
are expanded in a power series
\begin{equation}
  \mathbf{E}_{\rm T/R}(\mathbf{r}) = \sum_{n=0}^{\infty} \mathbf{E}^{(n)}_{\rm T/R} h^n. 
\end{equation}
Furthermore, all quantities in the Ewald-Oseen extinction theorem are expanded in a Taylor
series with respect to the small quantity $h f(x,y)$, giving
\begin{equation}
  \mathbf{E}_{\rm T}(\mathbf{r}') = \mathbf{E}_{\rm T}^{(0)}(x',y',z'=0) - h f(x',y')
                                  \frac{\partial\mathbf{E}_{\rm T}^{(0)}}{\partial z'}(x',y',z'=0) + \mathcal{O}(h^2)
\end{equation}
for the transmitted field and  
\begin{equation}
  g(\mathbf{r}-\mathbf{r}') = g(\mathbf{r}-\mathbf{r}')\Big|_{z'=0} -h f(x',y') \frac{\partial}
                            {\partial z'} g(\mathbf{r}-\mathbf{r}')\Big|_{z'=0} + \mathcal{O}(h^2)
\end{equation}
for the Green's function. The normal derivative times the surface element $\rd S'$ yields 
\begin{equation}
  \rd S' \frac{\partial}{\partial \mathbf{n}'} = - dx' dy'
                \left[\mathbf{e}_{\rm z} + \boldsymbol{\nabla}'_\parallel h f(x',y') \right]\cdot 
		\boldsymbol{\nabla}'\;. 
\end{equation}
Substituting these expressions into Eqs.~(\ref{Eq:Ewald1}) and (\ref{Eq:Ewald2}), and comparing
coefficients yields constitutive equations for the fields in different orders of $h$. 

Next, it is useful to expand the fields in plane waves according to 
\begin{equation}
  \begin{split}
    \mathbf{E}_{\rm T}(\mathbf{r}) &= \int \!\!\! \frac{\rd^2 \kappa}{(2\pi)^2}\, \re^{\ri (\boldsymbol{\kappa}\cdot\mathbf{x}
      +k_z z)} \boldsymbol{\mathcal{E}}_{\rm T}(\boldsymbol{\kappa})\\
    \mathbf{E}_{\rm R}(\mathbf{r}) &= \int \!\!\! \frac{\rd^2 \kappa}{(2\pi)^2}\, \re^{\ri (\boldsymbol{\kappa}\cdot\mathbf{x}
      -k_{z0} z)} \boldsymbol{\mathcal{E}}_{\rm R}(\boldsymbol{\kappa})\\
    \mathbf{E}_{\rm I}(\mathbf{r}) &= \int \!\!\! \frac{\rd^2 \kappa}{(2\pi)^2}\, \re^{\ri (\boldsymbol{\kappa}\cdot\mathbf{x}
      +k_{z0} z)} \boldsymbol{\mathcal{E}}_{\rm I}(\boldsymbol{\kappa}) 
  \end{split},
\end{equation}
and to utilize the Weyl expansion
\begin{equation}
  g(\mathbf{r}-\mathbf{r}') = \int \!\!\! \frac{\rd^2 \kappa}{(2 \pi)^2} \, \frac{\ri}{2 k_{z0}}
         \re^{\ri(\boldsymbol{\kappa}\cdot(\mathbf{x}-\mathbf{x}') + k_{z0} |z-z'|)}
\label{Eq:Weyl}
\end{equation}
for the Green's function, having introduced the notation $k_{z0}^2 = k_0^2 - \boldsymbol{\kappa}^2$,
$k_z^2 = k_0^2 \epsilon_r - \boldsymbol{\kappa}^2$, $\mathbf{x}:= (x,y)^t$, 
and $\boldsymbol{\kappa} = (k_x,k_y)^t$. It has to be
emphasized that the use of these expansions down to the surface relies on the Rayleigh 
hypothesis, as discussed in Ref.~\cite{M.Vesperinas06}.

With the help of the plane-wave expansions the relations for the Fourier components of the
reflected fields in terms of the incident field are easily calculated; for details we refer again
to the work by Agarwal~\cite{G.S.Agarwal1977}. For the zeroth-order field one obtains the expression 
\begin{equation}
  \boldsymbol{\mathcal{E}}_{\rm R}^{(0)}(\boldsymbol{\kappa}) = -\left[\frac{k_z-k_{z0}}{k_z+k_{z0}} \mathds{1}
        + \frac{2 k_{z0} (k_{z} - k_{z0})}
	 {k^2_0(k_{z0}\varepsilon_{r} + k_z)} \left(\boldsymbol{\kappa} - k_{z}k_{z0}\mathbf{e}_z\right)
	 \right]\boldsymbol{\mathcal{E}}_{\rm I}(\boldsymbol{\kappa}), 
\label{Eq:Ref0}
\end{equation}
and for the first-order field
\begin{equation}
  \boldsymbol{\mathcal{E}}_{\rm R}^{(1)}(\boldsymbol{\kappa}) = \ri (\varepsilon_{\rS} -1 ) \int\!\frac{\rd^2 \kappa'}
       {(2\pi)^2}\, F(\boldsymbol{\kappa} -\boldsymbol{\kappa}')\, \mathds{L}(\boldsymbol{\kappa})\, 
       \biggl[\frac{2 k_{z0}'}{k_z' + k_{z0}'} \mathds{1}  
       + \frac{2 k_{z0}' }{k_0^2 (k_{z0}' \epsilon_\rS + k_z')}\mathbf{K}_0'\otimes\mathbf{K}' \biggr]\boldsymbol{\mathcal{E}}_{I}(\boldsymbol{\kappa}'),  
\label{Eq:Ref1}
\end{equation}
with $\mathbf{K}' := (\boldsymbol{\kappa}',k_z')^t$, $\mathbf{K}_0' := (\boldsymbol{\kappa}',k_{z0}')^t$, and the dyadic operator  
\begin{equation}
  \begin{split}
  \mathds{L}(\boldsymbol{\kappa}) &:= \frac{1}{k_{z0}\varepsilon_\rS - k_z}\Big[(\kappa^2 +k_zk_{z0})
     (\mathds{1} - \mathbf{e}_z\otimes \mathbf{e}_z) - \boldsymbol{\kappa}\otimes\boldsymbol{\kappa}
      + k_z \mathbf{e}_z \otimes \boldsymbol{\kappa} + \varepsilon_r \kappa^2 \mathbf{e}_z \otimes
      \mathbf{e}_z\\
      &\qquad\qquad\qquad +\varepsilon_r k_{z0} \boldsymbol{\kappa}\otimes\mathbf{e}_z\Big].
  \end{split}
\end{equation}
Here $\otimes$ symbolizes the dyadic product of two vectors, and $\mathds{1}$ the unit dyad. Of course, also higher orders 
can be calculated within this approach~\cite{G.S.Agarwal1977}, but the higher-order contributions become increasingly 
cumbersome. Thus, this approach is particularly useful if meaningful results can already be obtained to first order. 
Hence, we restrict ourselves here to these first-order fields, and try to give approximate criteria justifying this
termination of the series later on.  
 
From Eq.~(\ref{Eq:E_field_greens_function}) it is clear that the Fourier component of the electric dyadic Green's 
function for the reflected fields given in Eqs.~(\ref{Eq:Ref0}) and (\ref{Eq:Ref1}) can be read off if 
we consider the incident electric fields 
$\boldsymbol{\mathcal{E}}_{\rm I}$ generated by a delta-like source current $\mathbf{j}_{e0}$
located at $\mathbf{r}''$ and put the Fourier component of the result into Eqs.~(\ref{Eq:Ref0}) and (\ref{Eq:Ref1}). 
According to Eq.~(\ref{Eq:E_field_greens_function}), this field can be stated as 
\begin{equation}
  \mathbf{E}_{\rm I}(\omega,\mathbf{r}) = \ri \omega \mu_0 \mathds{G}_0^{\rm E} (\mathbf{r},\mathbf{r}'')
                                   \cdot \mathbf{j}_{e0}, 
\label{Eq:incident_field0}
\end{equation}
where here $\mathds{G}_0^{\rm E}$ is the free dyadic Green's function, which can directly
be obtained from the relation 
\begin{equation}
   \mathds{G}_0^{\rm E}(\mathbf{r},\mathbf{r}'') = \left(\mathds{1} + \frac{\boldsymbol{\nabla}\otimes\boldsymbol{\nabla}}
                      {k_0^2}\right) g(\mathbf{r}-\mathbf{r}''). 
\end{equation}
Inserting the Weyl expansion from Eq.~(\ref{Eq:Weyl}) for the Green's function $g(\mathbf{r}-\mathbf{r}'')$,
such that $z''< z$, yields
\begin{equation}
   \mathds{G}_0^{\rm E}(\mathbf{r},\mathbf{r}'') = \int \frac{\rd^2 \kappa}{(2 \pi)^2} 
      \frac{\ri \re^{\ri(\boldsymbol{\kappa}\cdot(\mathbf{x}-\mathbf{x}'') + k_{z0}(z-z''))}}{2 k_{z0}}
      \Big(\mathbf{e}_{\perp}\otimes \mathbf{e}_{\perp} + \mathbf{e}_{\parallel}(-k_{z0}) \otimes
      \mathbf{e}_{\parallel}(-k_{z0})\Big). 
\label{Eq:free_Green}
\end{equation}
Here we have defined the unit vectors in vacuum for the TE- and TM-modes as
 \begin{equation}
   \mathbf{e}_{\perp} := \frac{1}{\kappa}\left(\,k_y\, , \,-k_x\, , \,0\, \right)^t
 \end{equation}
and
\begin{equation}
  \mathbf{e}_{\parallel}(k) := \frac{1}{\kappa k_0} 
                                \left(\,k_x k\, , \,k_y k\, , \,\kappa^2\,\right)^t.
\label{Eq:unit_tm}
\end{equation}
Using this expression for the free Green's function in Eq.~(\ref{Eq:incident_field0}) allows us to identify 
the Fourier component of the incident field, which is 
\begin{equation}
  \boldsymbol{\mathcal{E}}_{\rm I}(\boldsymbol{\kappa}) = -\omega \mu_0 \frac{\re^{-\ri(\boldsymbol{\kappa}\cdot
     \mathbf{x}'' + k_{z0} z'')}}{2 k_{z0}} \Big(\mathbf{e}_\perp\otimes\mathbf{e}_\perp 
      +\mathbf{e}_\parallel(-k_{z0})\otimes \mathbf{e}_\parallel(-k_{z0})\Big) \cdot \mathbf{j}_{e0}.
\label{Eq:incident_field}
\end{equation}
Now, it is a straightforward exercise to calculate the reflected fields. 
Substituting Eq.~(\ref{Eq:incident_field}) into Eq.~(\ref{Eq:Ref0}) gives the zeroth-order field 
\begin{equation}
  \boldsymbol{\mathcal{E}}_{\rm R}^{(0)}(\boldsymbol{\kappa}) = -\omega \mu_0 \frac{\re^{-\ri(\boldsymbol{\kappa}\cdot
     \mathbf{x}'' - k_{z0} z'')}}{2 k_{z0}} \Big( r_\perp \mathbf{e}_\perp\otimes\mathbf{e}_\perp 
     + r_{\parallel} \mathbf{e}_\parallel(k_{z0})\otimes \mathbf{e}_\parallel(-k_{z0}) \Big) \cdot\mathbf{j}_{e0}, 
\end{equation}
where we have introduced the usual Fresnel reflection coefficients for the TE- and TM-modes, defined as 
\begin{equation}
  r_\perp := \frac{k_{z0}-k_z}{k_{z0}+k_z}
\end{equation}
and
\begin{equation}
  r_\parallel := \frac{\varepsilon_{\rS} k_{z0}-k_z}{\varepsilon_{\rS}k_{z0}+k_z}.
\end{equation}
The first-order field can be calculated by substituting Eq.~(\ref{Eq:incident_field}) in Eq.~(\ref{Eq:Ref1}), 
giving
\begin{equation}
  \begin{split}
  \boldsymbol{\mathcal{E}}_{\rm R}^{(1)}(\boldsymbol{\kappa}) &= -\ri(\varepsilon_{\rS}-1) \int\! \frac{\rd^2 \kappa'}
    {(2\pi)^2}\, F(\boldsymbol{\kappa}-\boldsymbol{\kappa}')\, \mathds{L}(\boldsymbol{\kappa'})\, 
    \frac{\omega \mu_0}{2 k_{z0}}\, \re^{-\ri (\boldsymbol{\kappa}\cdot\mathbf{x}'' +k_{z0} z'')}\\
    & \qquad\qquad\qquad\qquad \times \left(t'_\perp \mathbf{e}'_\perp \otimes \mathbf{e}'_\perp + t'_{\parallel}
      \mathbf{e}'_\parallel(-k'_{z})\otimes \mathbf{e}_\parallel(-k'_{z})\right) \cdot \mathbf{j}_{e0},  
  \end{split}
\end{equation}
with the transmission coefficients for the TE- and the TM-modes defined as 
\begin{equation}
  t_\perp := \frac{2 k_{z0}}{k_{z0}+k_z}
\end{equation}
and 
\begin{equation}
  t_\parallel := \frac{2 k_{z0}}{\varepsilon_{\rS}k_{z0}+k_z}.
\end{equation}
Due to the definition of the vector $\mathbf{e}_\parallel$ for the TM-modes 
in Eq.~(\ref{Eq:unit_tm}) the TM-mode 
transmission coefficient defined here does not coincide with the standard formulation 
of the transmission coefficient. Finally, the reflected electric dyadic Green's function can be read off, 
giving the zeroth-order expression
\begin{equation}
  \mathds{G}_{\rm R}^{\rm E, 0}(\mathbf{r},\mathbf{r}'') = \int\! \frac{\rd^2 \kappa}{(2\pi)^2} 
  \frac{\ri \re^{\ri(\boldsymbol{\kappa}\cdot(\mathbf{x}-\mathbf{x}'') - k_{z0}(z+z''))}}{2 k_{z0}}
  \Big( r_\perp \mathbf{e}_\perp\otimes\mathbf{e}_\perp 
     + r_{\parallel} \mathbf{e}_\parallel(k_{z0})\otimes \mathbf{e}_\parallel(-k_{z0}) \Big),
\label{Eq:Green_ref_0}
\end{equation}
and the first-order expression
\begin{equation}
\begin{split}   
   \mathds{G}_{\rm R}^{\rm E, 1}(\mathbf{r},\mathbf{r}'') &= (1- \varepsilon_{\rS}) \int\!\frac{\rd^2
   \kappa}{(2\pi)^2} \int\! \frac{\rd^2 \kappa'}{(2\pi)^2} \,F(\boldsymbol{\kappa}-\boldsymbol{\kappa}')\,
   \mathds{L}(\boldsymbol{\kappa'})\,\frac{1}{2 k'_{z0}} \re^{\ri (\boldsymbol{\kappa} \cdot \mathbf{x} - k_{z0} z)}
   \re^{-\ri(\boldsymbol{\kappa'}\cdot\mathbf{x} +k'_{z0} z)}\\
    & \quad\quad \qquad \qquad \qquad \qquad \times \left(t'_\perp \mathbf{e}'_\perp \otimes \mathbf{e}'_\perp +
     t'_{\parallel} \mathbf{e}'_\parallel(-k'_{z})\otimes \mathbf{e}_\parallel(-k'_{z})\right).
\end{split}
\label{Eq:Green_ref_1}
\end{equation}
Since the magnetic dyadic Green's function is linked with the electric one by means of Eq.~(\ref{Eq:Green_electric_magnetic}),
we now have all ingredients to calculate the LDOS above a structured surface up to first order.

\section{Local density of states above a structure surface}
\label{Sec4}

The zeroth- and first-order Green's functions in Eqs.~(\ref{Eq:Green_ref_0}) and (\ref{Eq:Green_ref_1}) 
can now be used to calculate the electric and magnetic local density of states 
\begin{equation}
\begin{split}
  D^{\rE/\rH}(\omega,\mathbf{r}) &\approx D^{\rE/\rH}_0(\omega,\mathbf{r}) + D^{\rE/\rH}_1(\omega,\mathbf{r}) \\
                                 &:= \frac{\omega}{\pi c^2} \Im\,\Tr\,\mathds{G}^{\rE/\rH,0}_{R} 
                                    + h \frac{\omega}{\pi c^2} \Im\,\Tr\,\mathds{G}^{\rE/\rH,1}_{R}
\label{Eq:LDOS_01}
\end{split}
\end{equation}
by means of the equilibrium relations in Eqs.~(\ref{Eq:equi_electric_LDOS}) and (\ref{Eq:equi_magnetic_LDOS}).
As expected, and as a confirmation of the validity of the approach, 
to zeroth order we obtain the well-known expressions~\cite{JoulainEtAl03} 
\begin{equation}
\label{eq:DE0}
  D^{E}_0(\omega, a)=\frac{\omega}{4 \pi^2 c^2} \Im \, \int\limits_{0}^{\infty} \!
                       \text{d}k_x \, \frac{k_x \re^{-2\gamma a }}{2\gamma} \left( r_\perp
                       +\frac{2k_x^2-k_0^2}{k_0^2} r_\parallel \right)  
\end{equation}
and
\begin{equation}
  D^{H}_0(\omega, a)=\frac{\omega}{4 \pi^2 c^2} \Im \, \int\limits_{0}^{\infty} \!
                       \text{d}k_x \, \frac{k_x \re^{-2\gamma a }}{2\gamma} \left( r_\parallel
                       +\frac{2k_x^2-k_0^2}{k_0^2} r_\perp \right) 
\end{equation}
for the LDOS at the distance $a = -z$ above a semi infinite body,  
with $\gamma=\sqrt{k_x^2-k_0^2}$. The first-order contributions are
\begin{equation}
\begin{split}
  D^{E}_1 (\omega,a) =& h \Im \bigg[ \frac{1-\epsilon_\rS}{\omega \pi} \int\!
                        \frac{\text{d}^2 \kappa}{4 \pi^2} \,\int\!
                        \frac{\text{d}^2 \kappa'}{4 \pi^2} \,
                        F(\boldsymbol{\kappa}-\boldsymbol{\kappa'})\, \re^{\ri(\boldsymbol{\kappa}-\boldsymbol{\kappa'})\cdot \mathbf{x}}
                        \re^{\ri(k_{z0}+k_{z0}')a} \\
                        &\frac{t_\parallel'}{2 k_{z0}'}\frac{t_\parallel}{2 k_{z0}} \bigg\{ k_z
                        k_{z0}(\kappa'^2+k_z' k_{z0}')+k_z' k_{z0}'(\kappa^2+k_z k_{z0})\\ 
                        &-(\boldsymbol{\kappa} \cdot \boldsymbol{\kappa'})\big( k_z k_z' + \epsilon_S k_{z0} k_{z0}'
                        \big) + \epsilon_S \kappa^2 \kappa'^2+(\boldsymbol{\kappa} \cdot \boldsymbol{\kappa'} )^2
                        \bigg\} \bigg] \\
                      \equiv& h \Im \bigg[ \frac{1-\epsilon_\rS}{\omega \pi} \int\!
                        \frac{\text{d}^2 \kappa}{4 \pi^2} \,\int\!
                        \frac{\text{d}^2 \kappa'}{4 \pi^2} \,
                         F(\boldsymbol{\kappa}-\boldsymbol{\kappa'})\, \re^{\ri(\boldsymbol{\kappa}-\boldsymbol{\kappa'})\cdot \mathbf{x}}
                        I_\rE(\boldsymbol{\kappa},\boldsymbol{\kappa}') \bigg]
\end{split}
\label{eq:DE1}
\end{equation}
and
\begin{equation}
\begin{split}
D^{H}_1 (\omega,a) =& h \Im \bigg[ \frac{\epsilon_\rS-1}{\omega \pi} \int\!
                      \frac{\text{d}^2 \kappa}{4 \pi^2} \,\int\!
                      \frac{\text{d}^2 \kappa'}{4 \pi^2} \,
                      F(\boldsymbol{\kappa}-\boldsymbol{\kappa'}) \, \re^{\ri(\boldsymbol{\kappa}-\boldsymbol{\kappa'})\cdot \mathbf{x}}
                      \re^{\ri(k_{z0}+k_{z0}')a} \\
                     &\frac{t_\parallel'}{2 k_{z0}'}\frac{t_\parallel}{2 k_{z0}} \bigg\{ \bigg[
                      \frac{\kappa'^2 + k_z' k_{z0}'}{k_0^2} \big(k_{z0} k_{z0}'-\boldsymbol{\kappa} \cdot 
                      \boldsymbol{\kappa'}\big) + k_{z0}k_z' \bigg] \big( \kappa^2 + k_{z0} k_z \big)\\
                     &+k_{z0}' \big(k_z-k_{z0} \big) \frac{\big(\boldsymbol{\kappa} \times \boldsymbol{\kappa'}
                      \big)^2}{\kappa'^2}\frac{\kappa'^2 + k_{z0}'k_z'}{k_0^2}+k_z'\big( k_z -
                      k_{z0} \big) \frac{\big( \boldsymbol{\kappa} \cdot \boldsymbol{\kappa'}
                      \big)^2}{\kappa'^2}\\
                     &-\epsilon_S k_0^2 \boldsymbol{\kappa} \cdot \boldsymbol{\kappa'} \bigg\} \bigg] \\
                     \equiv& h  \Im \bigg[ \frac{\epsilon_\rS-1}{\omega \pi} \int\!
                      \frac{\text{d}^2 \kappa}{4 \pi^2} \,\int\!
                      \frac{\text{d}^2 \kappa'}{4 \pi^2} \,
                       F(\boldsymbol{\kappa}-\boldsymbol{\kappa'}) \,\re^{\ri(\boldsymbol{\kappa}-\boldsymbol{\kappa'})\cdot \mathbf{x}}
                      I_\rH(\boldsymbol{\kappa},\boldsymbol{\kappa}')  \bigg].
\end{split}
\label{eq:DH1}
\end{equation}
From these two equations~(\ref{eq:DE1}) and (\ref{eq:DH1}) it can be seen that in first order the electric
and magnetic local density of states 
are given by an integral of the Fourier transform 
\begin{equation}
  F(\boldsymbol{\kappa} - \boldsymbol{\kappa}') = \int \!\! \rd x  \int\!\! \rd y \, f(x,y) \,\re^{- \ri \mathbf{x} \cdot (\boldsymbol{\kappa} - \boldsymbol{\kappa}')} 
\end{equation}
of the profile $f(x,y)$, multiplied by the exponential factor $\exp(\ri (\boldsymbol{\kappa} - \boldsymbol{\kappa}') \cdot\mathbf{x} )$ 
and the function $I_\rE(\boldsymbol{\kappa},\boldsymbol{\kappa}')$ or $I_\rH(\boldsymbol{\kappa},\boldsymbol{\kappa}')$, respectively. 
Due to the factor $\exp(\ri(k_{z0}+k_{z0}')a)$, these two functions are exponentially damped for 
$\kappa$ and $\kappa'$ with modulus much greater than the inverse observation 
distance $a^{-1}$ in the near field, since in the evanescent regime $k_{z0} \approx \ri \kappa$ and $k_{z0}' \approx \ri \kappa'$. 
Therefore, the smaller the observation distance the more Fourier components 
contribute to the integral. Hence, from the structure of the first-order integrals it can be expected that for 
distances smaller than the characteristic width of the surface profile function the electric and magnetic local 
density of states resembles the surface profile. 

\begin{figure}[Hhbt]
  \centering
   \subfigure[Observation distance $a = 40 \, \text{nm}$]
     {\epsfig{file=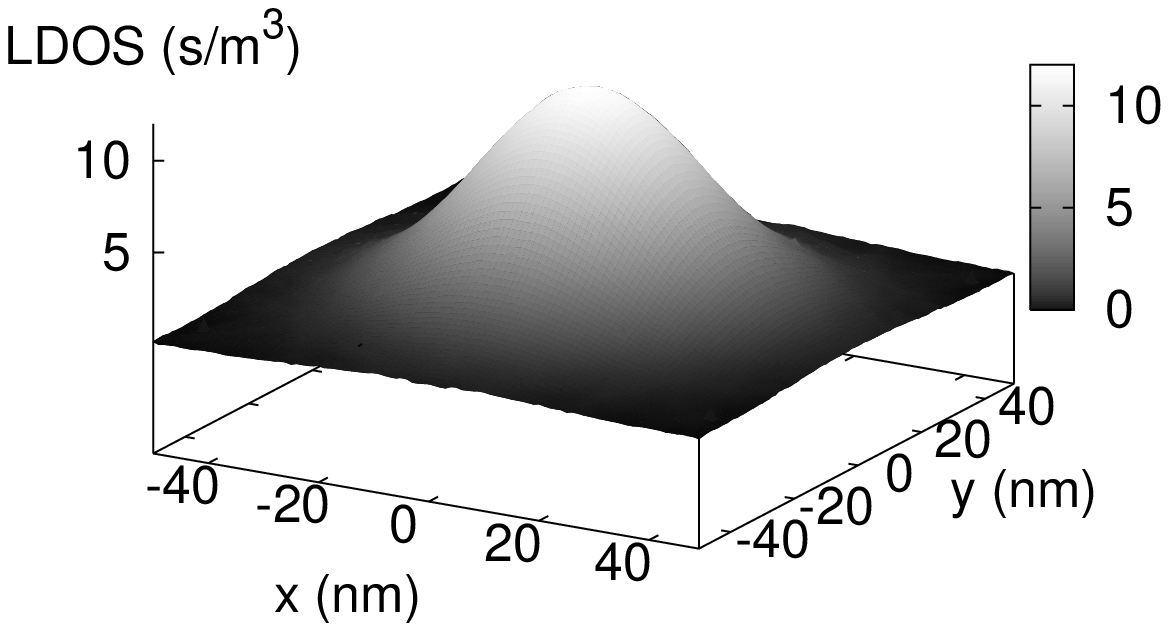, width = 0.48\textwidth}
     }
   \subfigure[Observation distance $a = 5.5 \, \text{nm}$]
    {\epsfig{file=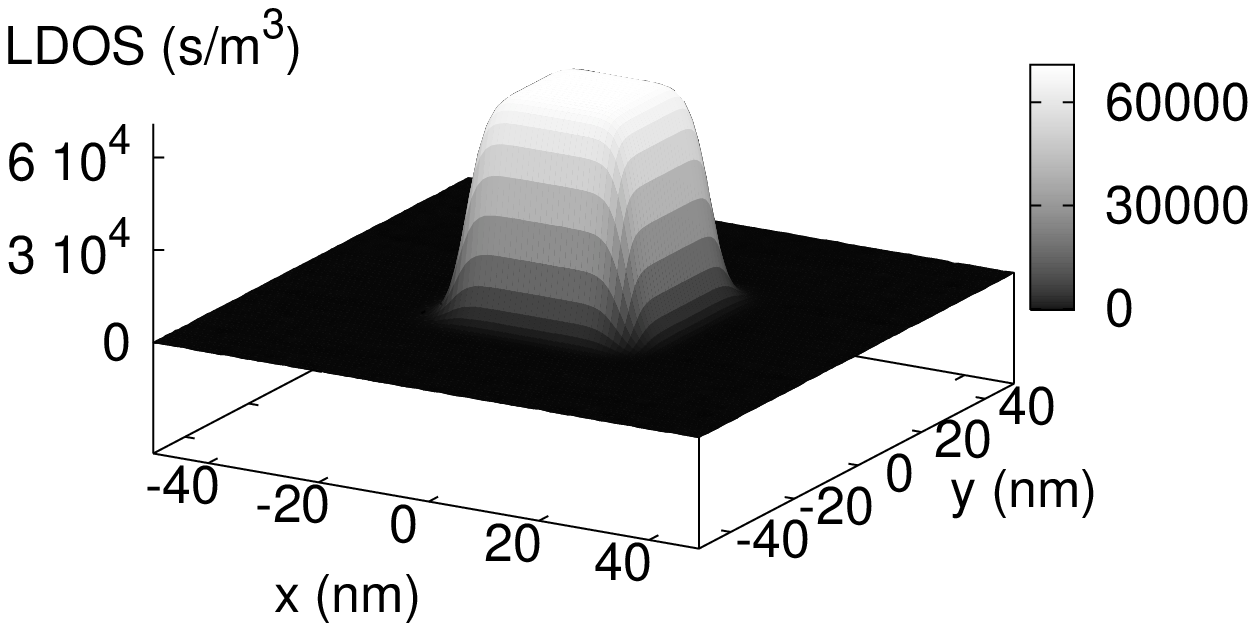, width = 0.48\textwidth}
    }

    \caption{Electric LDOS from eq.~(\ref{Eq:LDOS_01}) above a square pad with an edge length of $15 \, \text{nm}$
      and a height of $h = 5 \, \text{nm}$ for two different observation distances $a$ from the base plane. 
      The frequency considered is $\omega = 10^{14} \text{s}^{-1}$.}
    \label{fig:wuerfel}
\end{figure}

This behaviour is confirmed in Fig.~\ref{fig:wuerfel}, where the numerical evaluation of
the electric density of states from Eq.~(\ref{Eq:LDOS_01}) above a square pad with a height $h = 5\, {\rm nm}$ 
and width $w = 15 \,{\rm nm }$ on a plane surface is plotted. Here, we have used the 
frequency $\omega = 10^{14}\,{\rm s}^{-1}$ and the Drude permittivity $\epsilon_{\rS}$ of gold. 
Obviously, at an observation distance of $a = 40\, {\rm nm}$ the values of the local density of states resemble
a two dimensional bell-shaped function, whereas at an observation distance of $a = 5.5 \,{\rm nm}$,
being much smaller than the width of the square pad, the values of the local density 
mimic the underlying structure, apart from softened edges. 

\begin{figure}[Hhbt]
  \centering
    \epsfig{file=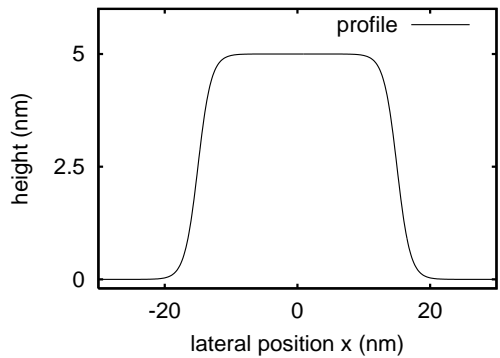, width = 0.47\textwidth}
  \caption{Plot of $h f(x,y)$ with the profile function defined by Eq.~(\ref{Eq:profile}) with 
           $d = 10^9 \, \text{m}^{-1}$, $w = 30 \, \text{nm}$, and $h = 5 \, \text{nm}$.}
  \label{Fig:profile}
\end{figure}

In the following, we discuss the magnitude of the contributions of the zeroth and first order of the electric and
magnetic local density of states. For that purpose, we assume an infinitely extended bar with a width $w$ 
on a plane surface modelled by the profile function
\begin{equation}
  f(x,y) = \frac{1}{\exp(d (|x|- \frac{w}{2}))+1}, 
  \label{Eq:profile}
\end{equation}
assuming $d = 10^9 \, \text{m}^{-1}$ and $w = 30 \, \text{nm}$. The height of the bar (see Fig.~\ref{Fig:profile}) 
is chosen to be $h = 5 \, \text{nm}$. 

\begin{figure}[Hhbt]
  \centering
  \subfigure[Electrical LDOS]{
    \epsfig{file=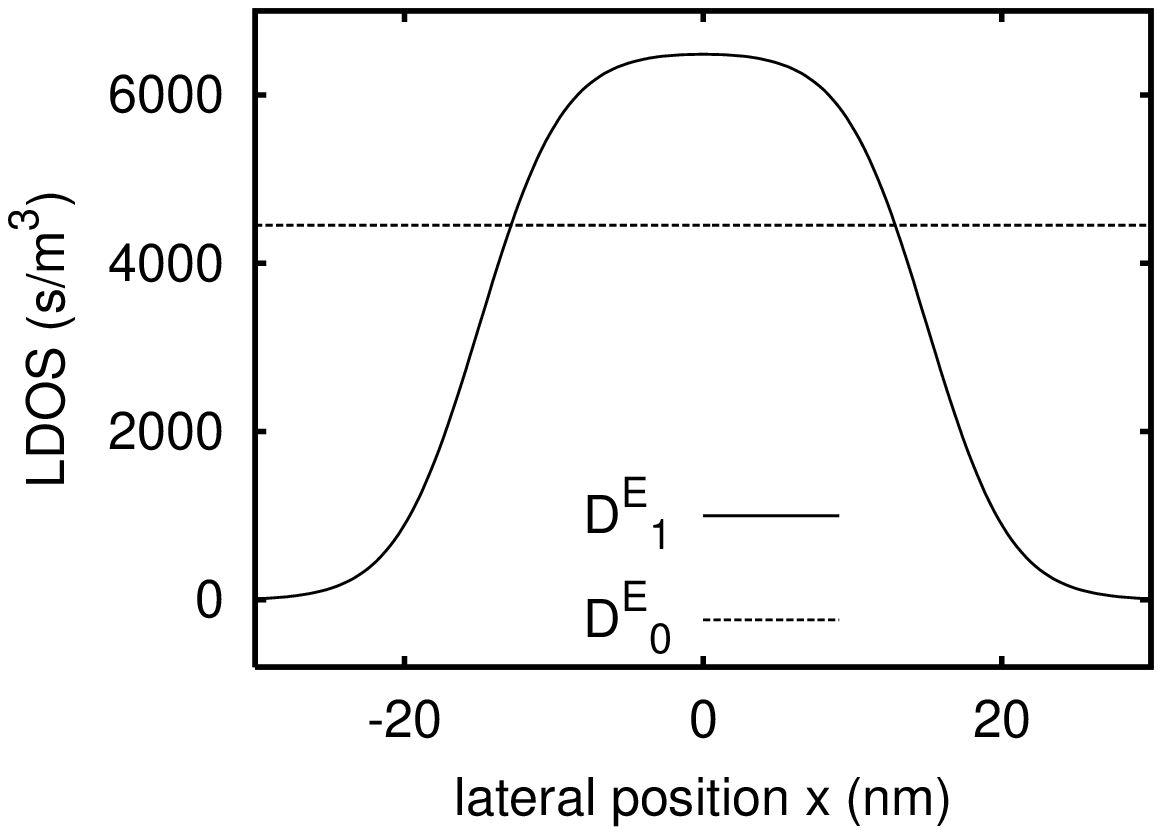, width = 0.47\textwidth}
  }
  \subfigure[Magnetical LDOS]{
    \epsfig{file=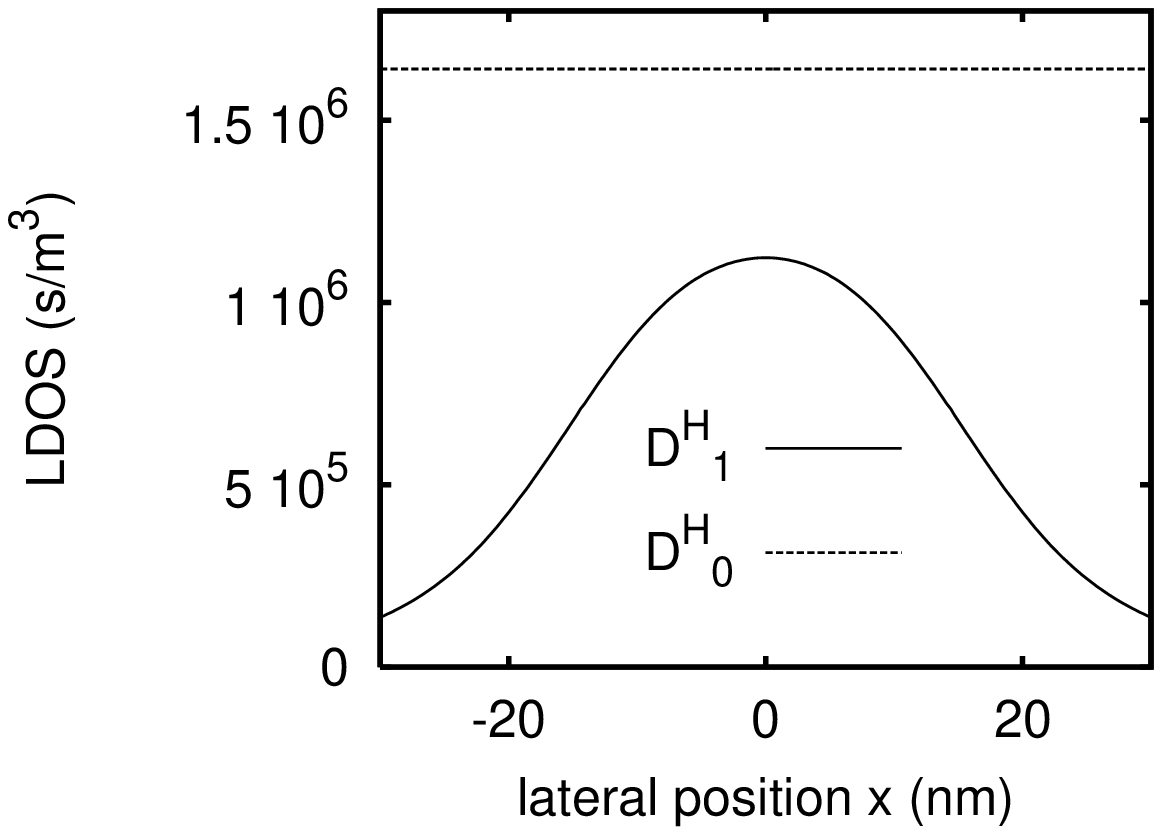, width = 0.47\textwidth}
  }
  
  \caption{LDOS of an gold half space structured by a gold bar (see Fig.~\ref{Fig:profile}). The 
  frequency used is $\omega=10^{14}\,{\rm s}^{-1}$, and the observation distance $10 \,
  \text{nm}$. Observe the different scales for the electric LDOS (a) and the magnetic LDOS (b)} 
  \label{fig:part_consth}
\end{figure}

The numerical values of the local density of states at a constant observation distance $a = 10\, {\rm nm}$ 
above the base plane are shown in Fig.~\ref{fig:part_consth},
using again the frequency $\omega = 10^{14} {\rm s}^{-1}$ and the Drude permittivity for 
gold. Firstly, one observes that the values of the magnetic LDOS are much greater than the
values of its electric counterpart, which is typical for metals, whereas for a polar dielectric 
bodies the electric LDOS usually gives greater values than the
magnetic one. Secondly, at the given observation distance we find values for the first-order contribution
to the electric LDOS, which are of the same order of magnitude
as the zeroth-order contribution. On the other hand, the values of the first-order magnetic LDOS are 
significantly smaller than the zeroth-order values for all lateral positions. Thirdly, the values of 
$D^{\rE}_1$ and $D^{\rH}_1$ give an equally good image of the bar on the plane surface,  
the width of the two bell-shaped curves being approximately the same. Therefore, in this case both first-order contributions give
qualitatively similar results.    

\begin{figure}[Hhbt]
  \centering
  \begin{minipage}[t]{0.9\textwidth}
    \epsfig{file=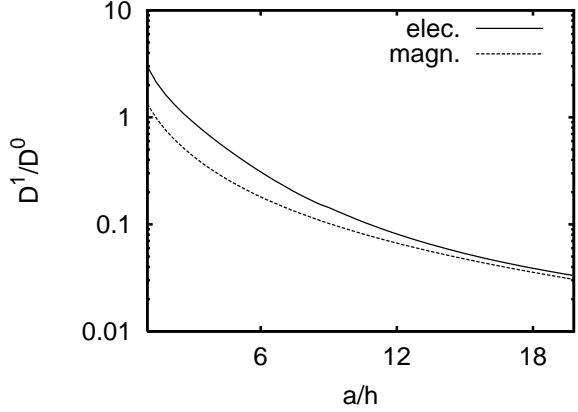, width = 0.6\textwidth}
    \caption{Dependence of the ratio of the first and zeroth order of the LDOS
    on the observation distance $a$ calculated above the profile shown in Fig.~\ref{Fig:profile}
    for $x = 0$ and frequency $\omega= 10^{14} \, \text{s}^{-1}$.} 
    \label{fig:z-spekto}
  \end{minipage}
\end{figure}

In Fig.~\ref{fig:z-spekto} we plot
the ratio of the first- and zeroth-order LDOS $D_1/D_0$ for the same surface profile directly above 
the bar, i.e.\ at $x = 0$, for observation distances ranging from $6\, {\rm nm}$ to $100 \,{\rm nm}$. 
As expected, this ratio increases for decreasing observation distance, suggesting that for small
distances higher-order terms have to be considered. These higher-order terms can in principle 
be calculated by an iterative scheme deduced by Greffet~\cite{Greffet1988}. We 
suggest that it might be sufficient to consider only $D_0$ and $D_1$ as long as the ratio $D_1/D_0$
does not exceed the ten percent level. This means that the numerical results depicted in Figs.~\ref{fig:part_consth} a)
and b) refer to distances where higher-order terms should be taken into account. 
Furthermore, it is evident from Fig.~\ref{fig:z-spekto} that for all distances the ratio of the leading 
two contributions to the magnetic LDOS gives much smaller values than the corresponding ratio 
for the electric LDOS. Therefore, it can be concluded 
that the approximation committed when considering only the zeroth- and first-order terms holds 
for the magnetic part for much smaller distances than for the electric part. This also means 
that the underlying structure becomes important in the electric LDOS for much greater distances 
than in the magnetic LDOS.     
 
\begin{figure}[Hhbt]
  \centering
  \begin{minipage}[t]{0.9\textwidth}
    \epsfig{file=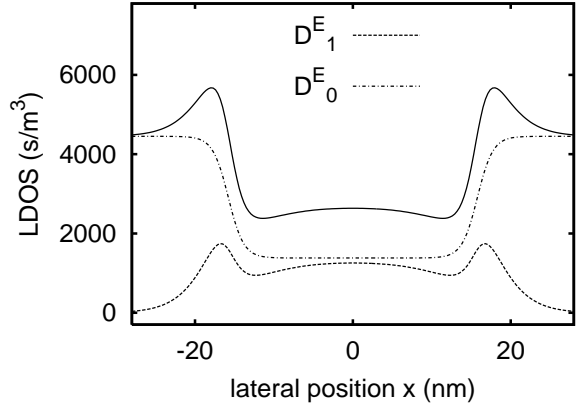, width = 0.6\textwidth}
    \caption{Electric local density of states ($D^\rE_0$: dash-dotted line; $D^\rE_1$: dashed line; 
             $D^\rE_0 + D^\rE_1$: solid line) above the structured surface as
             depicted in Fig.~\ref{Fig:profile} at a frequency of 
             $\omega=10^{14} \, \text{s}^{-1}$ and a constant separation of $10 \,
             \text{nm}$ between observation point and surface.}
    \label{fig:part_constd}
  \end{minipage}
\end{figure}

So far we have determined the LDOS at a constant height above the base plane of the structured
surface. A very common operation mode in near-field microscopy is the constant-distance mode, where the 
separation between the tip and the individual features of the sample is kept
constant~\cite{WishnathEtAl08}. In order to calculate the LDOS relevant for this mode, we use the observation
distance $a+h f(x,y)$ instead of $a = const.$ in Eq.~(\ref{Eq:LDOS_01}). 
Fig.~\ref{fig:part_constd} shows the LDOS obtained for this 
constant-distance mode above the structured half space depicted in Fig.~\ref{Fig:profile}.

In the constant-height mode (see Fig.~\ref{fig:part_consth} (a) and (b)) the first-order term
$D_1$ gives a rough image of the underlying surface structure, while the
zeroth-order term $D_0$ is constant. However, in the constant-distance 
mode (see Fig.~\ref{fig:part_constd}) the qualitative 
behaviour of the two contributions is more complex, due to the variation of the observation distance.
Here, the zeroth-order term $D_0$ coincides more or less with the inverse of the underlying structure,
and the first-order term $D_1$ gives 
relatively large values near the edge of the bar, due to the variation in observation distance. Therefore, 
at least two regimes of observation distances can be distinguished: At large distances, where $D_1/D_0$ is small, the LDOS is dominated
by the zeroth-order term, giving values which coincide approximately with the inverse of the surface
structure. At distances where at least the first-order term has to be taken into account, the variation in distance 
leads to a rather complex pattern of the LDOS.

\section{Consequences for the near-field scanning thermal microscope}
\label{Sec5}

Up to this point, only the LDOS at a frequency corresponding to $300 \, \text{K}$ have been 
calculated. Before we discuss some numerical results for the full near-field radiative 
heat transfer between a metallic nano sphere and a structured surface, we  
emphasize the restrictions of the model: Firstly, due to the dipole approximation this
model is only valid for distances $a$ significantly greater than the radius of the sphere $R$. 
Secondly, for distances and sphere radii smaller than the mean free path of the conduction 
electrons, nonlocal and quantum mechanical effects become important and have  
to be implemented. Thirdly, due to the perturbative approach the results apply under the condition 
that the height $h$ of the profile is the smallest length scale, i.e.\ $h \ll \min\{a,\lambda_{\rm th},w\}$
with the width of the surface structure $w$ and the thermal wavelength $\lambda_{\rm th} \approx \hbar \beta c$.
In particular this means that the pertubative approach is not valid for 
distances $a \ll R$; in this limit the ``proximity approximation'' can be 
utilised~\cite{VolokitinPersson2007}. Nonetheless, it is reasonable to explore the perturbative 
predictions in some detail. 

\begin{figure}[Hhbt]
  \centering
  \subfigure[Heat transfer in the constant-height mode.]
  {\epsfig{file=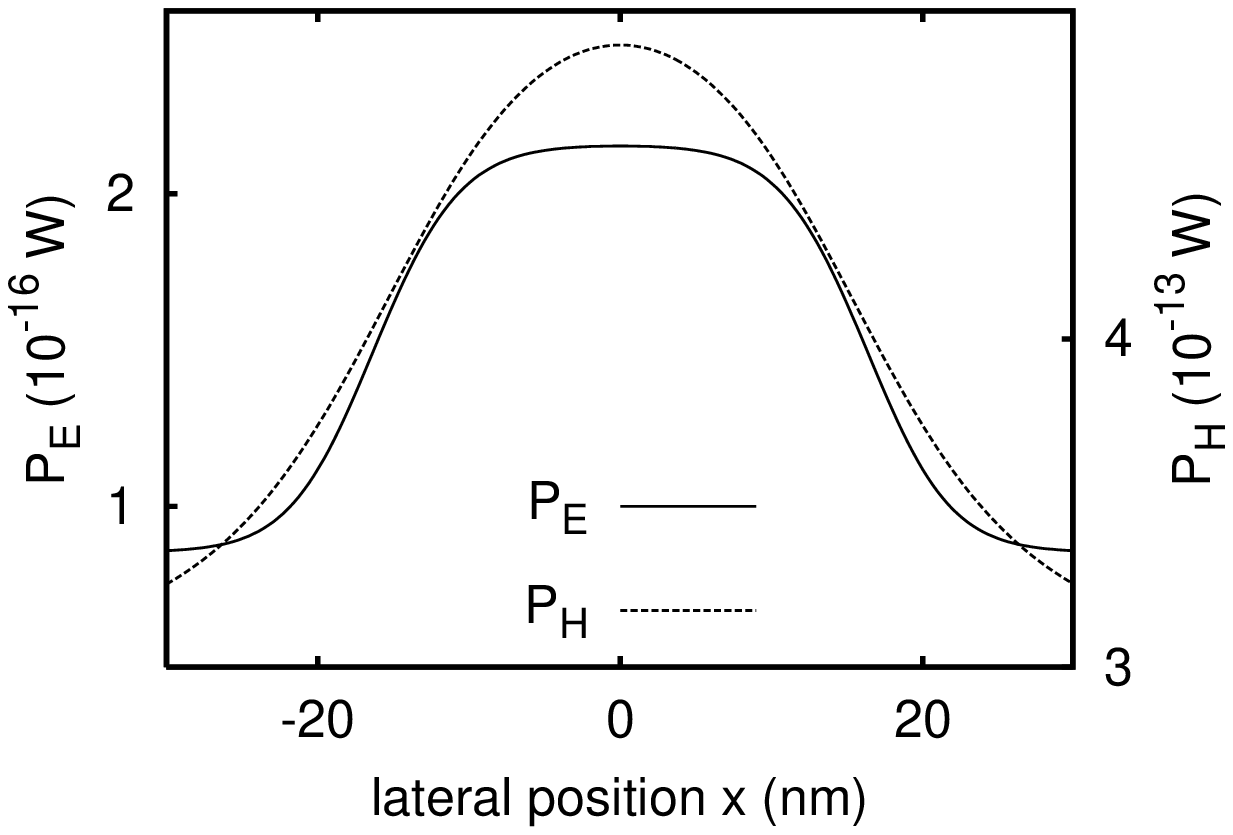, width = 0.48\textwidth}
  }
  \subfigure[Heat transfer in the constant-distance mode.]
  {\epsfig{file=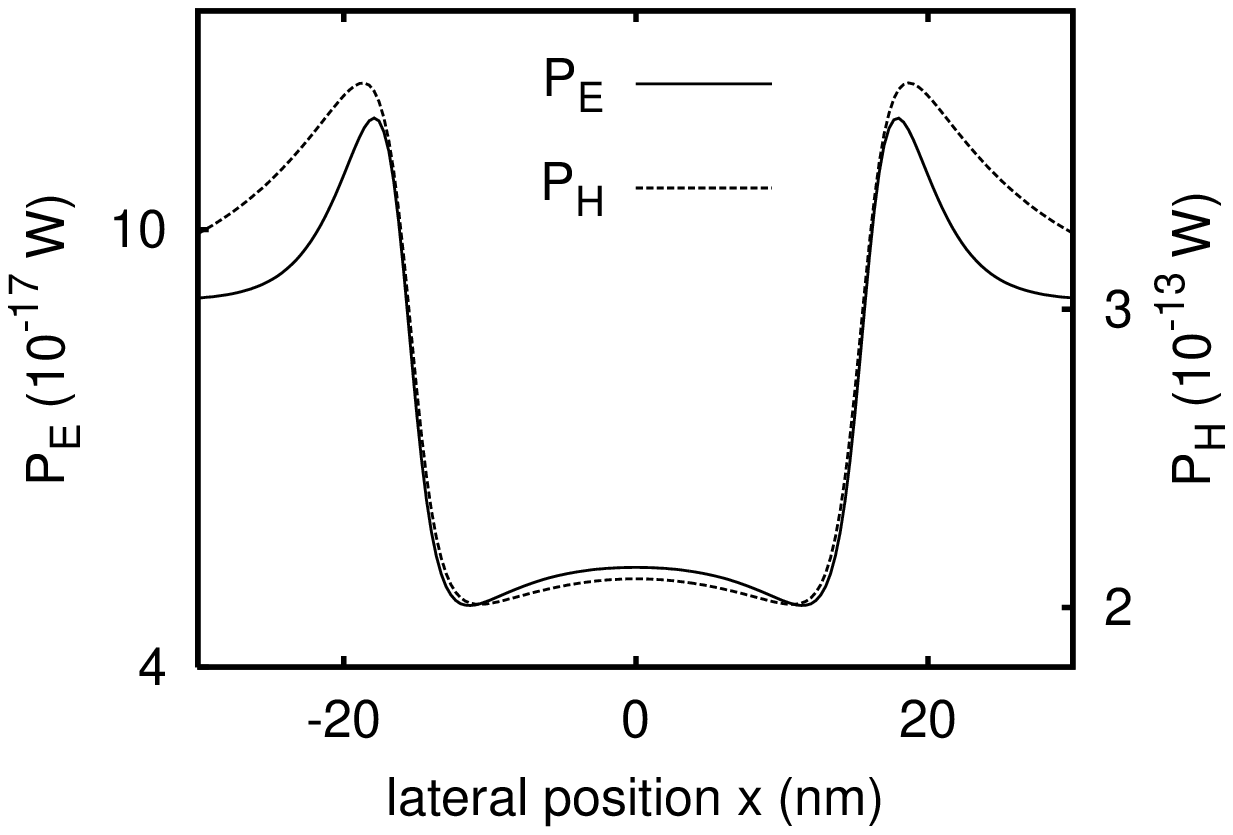, width = 0.48\textwidth}
  }
  \caption{Near-field radiative heat transfer between the structured sample 
      depicted in Fig.~\ref{Fig:profile} ($T_\rS = 100 \, \text{K}$) and
      a spherical probe ($T_\rP = 300 \, \text{K}$) with a radius of 
      $10 \, \text{nm}$. The height and the separation was $10 \, \text{nm}$ in each case.}
    \label{fig:p}
\end{figure}

To this end, we evaluate the near-field radiative heat transfer between a gold nanoparticle 
and a gold sample as given by Eq.~(\ref{Eq:DipolModel}), using the polarisabilities 
of the sphere~(\ref{Eq:alpha_E}) and (\ref{Eq:alpha_H}) with the
Drude permittivities $\epsilon_\rP$ and $\epsilon_\rS$ for gold. As surface
profile we employ the infinitely extended bar from Eq.~(\ref{Eq:profile}) with parameters as 
in Fig.~\ref{Fig:profile}. In order to relate the results for the energy flow $P$ to the numerical results for the LDOS we evaluate 
Eq.~(\ref{Eq:DipolModel}) for a constant height,  $a = 10\, {\rm nm}$, and a 
constant distance, $a = 10\, {\rm nm} + h f(x,y)$, where
a nanoparticle radius of $R = 10 \,{\rm nm}$ has been chosen. As shown in Fig.~\ref{fig:p}, in both cases 
the near-field radiative heat transfer is dominated by the magnetic contribution, this 
being a typical feature for good metals like gold (at the given distance). Apart from 
this the curves displayed in Fig.~\ref{fig:p} reflect the corresponding plots of the LDOS 
at a frequency near the thermal frequency $\omega_{\rm th}$. 
This is a consequence of Eq.~(\ref{Eq:DipolModel}), since the main contributions to the frequency
integral stem from frequencies near the thermal frequency as long as there are no resonances 
in the thermally accessible frequency regime. We also implemented the ``proximity approximation''
for the given geometry numerically (with the same parameters as in Fig.~\ref{fig:p}), obtaining qualitatively 
similar results for the near-field radiative heat transfer as those given by the dipole model.

Hence, the conclusions deduced in the last section for the local density of states can be transfered to 
the near-field radiative heat transfer between a sphere and a structured surface, modelling a NSThM tip
and a structured sample. Thus, for example we conclude that the lateral resolution of a NSThM
increases for decreasing distances, as long as this microscope can be described within the dipole model. 
The lateral resolution is comparably good for the electric part $P_\rE$ and the magnetic part $P_\rH$, with
the topology of the surface profile being well resolvable for distances much smaller than the width of the surface pattern. 
On the other hand, for non-metallic materials, i.e.\ in a situation with $P_\rE \gg P_\rH$, the underlying structure 
becomes important for much greater distances than for metallic materials, as can be concluded from Fig.~\ref{fig:z-spekto}.

Furthermore, a NSThM gives very different images of the underlying structure when either the constant-height or
the constant-distance mode is used. In particular, for distances where $D_0 \gg D_1$ the values for $P$ should 
resemble the inverse surface structure in constant-distance mode, whereas for distances 
where $D_1$ becomes important, the measured signal can be rather complex in that mode. 
On the other hand, in constant-height mode the signal should approximately resemble the 
topology of the underlying structure at distances where $D_1$ becomes important. These are strong predictions
that are amenable to immediate verification in current experiments~\cite{WishnathEtAl08}, and may help to correctly
interprete the various types of signals obtainable with a NSThM~\cite{Preprint08}.

\acknowledgments
This work was supported in part by the Deutsche Forschungsgemeinschaft through grant No.\ KI 438/8-1.
We also thank M.~Holthaus for useful comments on the text.

\end{document}